\documentclass[reprint,prl,aps,amsmath,amssymb]{revtex4-2}

\usepackage{graphicx}
\usepackage{dcolumn}
\usepackage{bm}
\usepackage{hyperref}
\usepackage{appendix}

\begin{document}

\title{Optimal Control of a Mesoscopic Information Engine}
\author{Emanuele Panizon}
\affiliation{Area Science Park, Localit\`{a} Padriciano 99, 34149 Trieste, Italy}
\date{\today}

\begin{abstract}
We analytically solve the finite-time control problem of driving an overdamped particle via an optical trap under costly measurement. By formulating this mesoscopic information engine within the Partially Observable Markov Decision Process (POMDP) framework, we demonstrate that the underlying Linear-Quadratic-Gaussian (LQG) dynamics decouple the optimal measurement and driving protocols. We derive the optimal feedback control law for the trap placement, which recovers the discontinuous Schmiedl-Seifert driving protocol in the open-loop limit and extends it to any measurement scheduling. For a costly, binary (on/off) sensor, we evaluate the optimal measurement protocol and derive physical bounds on the maximum net gain that can be extracted from thermal fluctuations. We show the emergence of deadline-induced blindness, a phenomenon where all measurements cease as the deadline approaches regardless of their cost. Taking the infinite-horizon limit, we find the exact periodic measurement schedules for the steady state as a function of the measurement cost $C$ and derive the macroscopic velocity envelopes beyond which viscous drag forces the engine into a net-dissipative regime. Finally, we generalize the results to a variable-precision sensor. 
\end{abstract}

\maketitle

\textbf{Introduction.} 

Thermal fluctuations dominate mesoscopic dynamics. Stochastic thermodynamics and fluctuation theorems \cite{jarzynski1997nonequilibrium,crooks1999entropy,seifert2012stochastic} formalized the physical relations governing non-equilibrium trajectories, expanding the second law from macroscopic averages to individual microscopic realizations. 

When extended to systems with feedback, this framework established how information acquisition can be used to extract work from fluctuations\cite{sagawa2010generalized,bauer2012efficiency,sagawa2012,ehrich2023energy}. This concept, originating with Maxwell's Demon \cite{maxwell1871theory,szilard1929entropieverminderung} and bounded by the thermodynamics of information \cite{landauer1961irreversibility,parrondo2015thermodynamics}, has recently been realized in experimental laboratories. Specifically, colloidal particles manipulated by harmonic optical traps serve as prototypical information engines, where measurement-driven feedback loops have been successfully used to convert spatial fluctuations into directed motion or stored energy \cite{toyabe2010experimental,berut2012experimental,martinez2017colloidal}. Heuristic continuous-feedback ratchets, for example, perform well empirically in both thermal \cite{paneru2020colloidal,saha2021maximizing} and non-equilibrium baths \cite{saha2023information}.

However, how to construct \textit{optimal} feedback loops is still an open problem, even when measurement costs are not considered explicitly. Exact analytical solutions are sparse, as they typically require solving complex continuous Hamilton-Jacobi-Bellman integro-differential equations \cite{bensoussan1992stochastic,alvarado2026optimal}. Recent advances in optimal transport theory for fully observable continuous Langevin systems \cite{aurell2011optimal,kamijima2025optimal} and generalized geometric frameworks \cite{mohite2025generalized} have established discontinuous endpoint jumps \cite{blaber2021steps} as a generic physical mechanism to minimize driving dissipation. 

For the specific case of finite-time operation of optical traps, analytical solutions focus on the optimization of the physical control law \textit{conditional} on a fixed measurement architecture. Current models have described analytically the optimal trap protocol in the Schmiedl-Seifert regime (no feedback) \cite{schmiedl2007optimal}, the Abreu-Seifert limit (a single initial observation) \cite{abreu2011extracting}, under periodic measurement loops \cite{garcia2025optimal,schuttler2025active}, and (approximately) for continuous, perfect measurement\cite{whitelam2023demon}.

Since the literature optimizes the trap trajectory in isolation—rather than treating physical control and measurement scheduling as a single conjoint problem—the exact finite-time dynamics of costly measurements remain, to the best of our knowledge, unsolved.

In this work, we formulate the problem of an overdamped particle in a harmonic trap in discrete time in the framework of a Partially Observed Markov Decision Process\cite{bechhoefer2015hidden,biehl2022interpreting}. As such, we exploit Linear-Quadratic-Gaussian (LQG) control theory \cite{meier1967optimal,anderson2007optimal,bechhoefer2021control} to bypass the complex integro-differential equations of standard continuous-time partial observability. We show that the structure of thermodynamic work in a harmonic potential allows the measurement policy and the physical control law to decouple into simple algebraic equations. This decoupling yields a closed-form mapping of the demon’s complete thermodynamic phase space.

We derive the analytical control law $\lambda_k^*$ for the optimal trap trajectory, demonstrating how discrete finite-time feedback recovers and extends the continuous limits previously found in \cite{abreu2011extracting,abreu2011extracting}. 
By studying the extraction power limits of this information engine under two formulations of measurement costs, we map the thermodynamic boundaries of the system. Beyond general cost thresholds - beyond which observations can never be profitable - we show the existence of ``deadline blindness'' for finite-time operations. This condition is triggered when the threshold for profitable information, which diverges close to the terminal point, crosses the thermal equilibrium variance. We further establish the optimal periodic measurement schedule at the steady state for binary sensors and derive the macroscopic velocity boundaries that bankrupt the active engine via viscous drag. Finally, we  define the optimal precision acquisition rate for variable-precision sensors that transforms the discrete demon into a continuous ``information thermostat''. 

\vspace{12pt}
\textbf{Model and POMDP Formulation.} We consider an overdamped Brownian particle in a one-dimensional harmonic trap $U(x, \lambda) = \frac{1}{2}\kappa(x-\lambda)^2$ with stiffness $\kappa$, immersed in a heat bath at temperature $T$ and with friction coefficient $\gamma$. The objective is to steer the trap center from an initial position $\lambda_0 = 0$ to a target $\lambda_f$ over a finite time horizon $t_f$, minimizing the expected thermodynamic work.
Time $t$ is discretized into discrete steps $k\Delta t$, $k \in \{0,1,\dots N\}$. We cast this as a Partially Observable Markov Decision Process (POMDP)\cite{astrom1965optimal,kaelbling1998planning} where the hidden system state is the particle's true position $x_k$. Because the agent cannot observe $x_k$ for free, it tracks a belief state consisting of a Gaussian distribution, parameterized by its mean $\mu_k$ and variance $\Sigma_k$, augmented by the historical trap position $\lambda_{k-1}$.

At each time step, the temporal sequence of actions unfolds as follows: first, the agent decides 
how to actively measure its state, possibly paying a cost $C_k=C(\Sigma_k^+,\Sigma_k^-)$ to update its belief from a pre-measurement state $(\mu_k^-, \Sigma_k^-)$ into $(\mu_k^+, \Sigma_k^+)$. Two different possibilities for the sensors will be discussed, both a binary, on/off, perfect sensor and a variable-sensitivity one.
After the measurement, and exploiting the updated belief, the agent executes a physical control action by setting the new trap position $\lambda_k$. The step-cost associated with this physical action is the expected thermodynamic work done \textit{on} the system during the instantaneous trap shift:

\begin{equation}\mathbb{E}[W_k] = -\kappa (\lambda_k - \lambda_{k-1}) \left( \mu_k^+ - \frac{\lambda_k + \lambda_{k-1}}{2} \right).
\end{equation}

\begin{figure}
    \centering
    \includegraphics[width=0.9\linewidth]{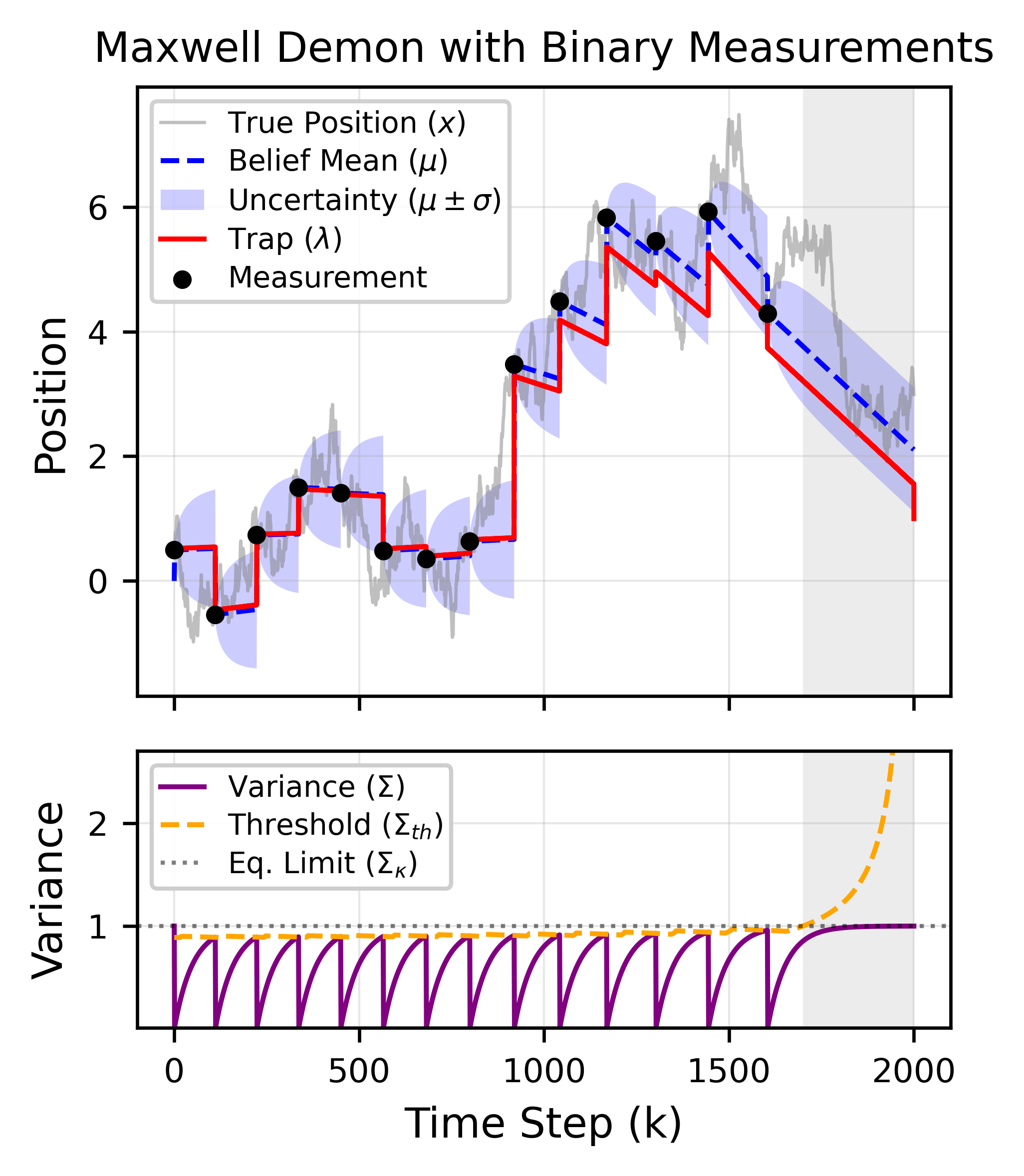}
\caption{\textbf{Optimal finite-time control dynamics of the information engine with binary measurements.} (Top) Simulated spatial trajectory illustrating the true particle position ($x$, gray), the inferred belief mean ($\mu$, blue dashed) with $1\sigma$ uncertainty bounds (blue shaded), and the optimal deterministic trap placement ($\lambda$, red). Black markers denote discrete measurement events. (Bottom) Corresponding temporal evolution of the belief variance ($\Sigma$, purple). Thermal diffusion drives the uncertainty upward until it intersects the dynamically computed optimal threshold ($\Sigma_{th}(n)$, orange dashed), triggering a measurement that collapses the variance. As the deadline approaches ($t \to t_f$), the required threshold for profitable measurement rigorously diverges, demonstrating the onset of finite-time deadline blindness, here shown in grey when $\Sigma_{th}(n) > \Sigma_\kappa$. Simulation parameters: $k_B T=1$, $\kappa=1$, $\gamma=1$, $\Delta t=0.01$, $t_f=20$, $C=0.6 \, C_{max}$.}
\label{fig:trajectory}
\end{figure} 

Following standard thermodynamic convention, $\mathbb{E}[W_k] < 0$ denotes net work extracted from the thermal bath. The quadratic state terms cancel out, making the expected physical work independent of the posterior variance $\Sigma_k^+$. Finally, the physical system evolves and the belief is modified following to the Ornstein-Uhlenbeck updates\cite{uhlenbeck1930theory,gillespie1996exact}:
\begin{align}
\mu_{k+1} &= \mu_k^+ \alpha + \lambda_k (1 - \alpha), \\
\Sigma_{k+1} &= \Sigma_k^+ \alpha^2 + \Sigma_{eq},
\end{align}
where $\alpha = \exp(-\kappa \Delta t / \gamma)$ and the equilibrium variance is $\Sigma_{eq} = (k_B T / \kappa)(1 - \alpha^2) = \Sigma_\kappa(1-\alpha^2)$.

\vspace{12pt}
\textbf{The Bellman Optimality Equation.} 
The objective of the agent is to determine the feedback policy that minimizes the total expected cost over the finite horizon $\mathcal{J}$. The total cost accumulates the thermodynamic work done on the system ($W_k$) and the energy spent to measure ($C_k$) at all time step $k$:

\begin{equation}
\mathcal{J}=\mathbb{E}\left[\sum_{k=0}^N (W_k + C_k)\right].
\end{equation}

We can formalize this minimization objective through the optimal cost function $V_n(\mu^-, \Sigma^-, \lambda)$, which evaluates the minimum (optimal) expected cost-to-go for the remaining steps $n=N-k$, conditioned on the current prior spatial estimate $\mu^-$, the prior uncertainty $\Sigma^-$, and the trap position $\lambda$.

The optimal joint protocol of measurement and physical trap displacement satisfies the recursive Bellman optimality equation\cite{bellman1957dynamic,bertsekas2012dynamic}:
\begin{equation}\label{eq:bellman_main}
\begin{split}
V_n(\mu^-, \Sigma^-, \lambda) = \min_{\Sigma^+} \Big\{ C(\Sigma^+, \Sigma^-) + \\
\mathbb{E}_{\mu^+} \left[ \min_{\lambda'} \left\{ \mathbb{E}_{x}[W] + V_{n-1}(\mu^{\prime-}, \Sigma^{\prime-}, \lambda') \right\} \right] \Big\},
\end{split}
\end{equation}
where $(\mu^{\prime-}, \Sigma^{\prime-})$ denote the prior belief state at the subsequent time step.

Since the continuous dynamics are governed by linear Langevin equations and the thermodynamic work is quadratic in the spatial coordinates, the system operates in the Linear-Quadratic-Gaussian (LQG) regime\cite{simon1956dynamic,bar1974dual,bechhoefer2021control}. The Separation Principle 
therefore guarantees that the informational and spatial controls can be separated. As shown in the Supplemental Material (SM), the value function $V_n$ decouples into two independent components concerning either the spatial or the informational degrees of freedom:
\begin{equation}\label{eq:decoupling_main}
V_n(\mu^-, \Sigma^-, \lambda) = J_n(\mu^-, \lambda) + g_n(\Sigma^-).
\end{equation}
Here, $J_n$ is the deterministic physical cost depending only on the spatial coordinates, and $g_n$ is the informational cost depending only on the estimation variance.  This allows us to separately tackle the optimal trap control and the optimal measurement scheduling. 

\vspace{12pt}
\textbf{The Riccati Recurrence.}
In standard discrete-time LQG control, finding the optimal physical policy reduces to propagating a quadratic value function backward in time (hence the use of the index $n$). This requires solving a discrete-time recurrence\cite{anderson2007optimal,bertsekas2012dynamic} for a cost matrix $\mathbf{P}_n$, called the Riccati equation. Given the augmented state space, $(\mu, \lambda)$, this is a $2 \times 2$ matrix. However, the thermodynamic work in a one-dimensional harmonic potential separates into terms dependent on either the current action $\lambda_k$ or the previous trap position $\lambda_{k-1}$. The absence of cross-terms means that the optimal feedback law $\lambda_k^*$ is independent of the previous trap placement.

This decoupling, as shown in the SM, reduces the matrix Riccati recurrence to a single scalar sequence for the spatial cost coefficient $P_n$, where $n = N - k$ is the number of remaining steps. The value function's stationary point defines the optimal feedback gain $K_n = \alpha P_{n-1} / (P_{n-1}(1-\alpha) + 2\kappa)$ and, in turn, the control law $\lambda_k^* = K_n \mu_k^+ + (1-K_n)\lambda_f$. 
Substituting the closed-form solution for $P_n$ (derived in the SM),
\begin{equation}\label{eq:p_exact}
P_n = -\kappa \left[ \frac{n(1-\alpha)}{1+\alpha+n(1-\alpha)}\right],
\end{equation}
returns the the explicit optical trap placement:
\begin{equation}\label{eq:lambda_exact}
\lambda_k^* = \mu_k^+ + \frac{1}{n(1-\alpha) + 1 + \alpha} (\lambda_f - \mu_k^+).
\end{equation}

Physically, the optimal control linearly interpolates between the current belief of the particle position $\mu_k^+$ and the final target $\lambda_f$. Far from the deadline ($n \Delta t \gg \gamma/\kappa$), the time-dependent weight approaches zero, and the trap closely follows the inferred position ($\lambda_k^* \approx \mu_k^+$) to optimally exploit thermal fluctuations. As the deadline approaches ($n \to 1$), the weighting shifts to drive the system toward the target $\lambda_f$. 


\vspace{12pt}
\textbf{The Continuous-Time Limit.} 
To demonstrate that our finite-time discrete control reproduces and extends established continuous-time frameworks, we take the continuous limit $\Delta t \rightarrow 0$. Expanding the relaxation parameter to first order, $1-\alpha \approx \Delta t/\tau$ (where $\tau = \gamma/\kappa$), we obtain the continuous-time limit for Eq.~\ref{eq:lambda_exact}:

\begin{equation}
\lambda^*(t) = \mu(t) + \frac{\tau}{(t_f - t) + 2\tau} (\lambda_f - \mu(t))
\end{equation}

It is important to recall that by the separation principle of LQG control, the optimal trap policy $\lambda^*(t)$ is independent of the measurement architecture; it is defined entirely by the instantaneous state estimate $\mu(t)$. As a consequence, this optimal law is \textit{always optimal}: Different costs and definitions of the sensors will result in different optimal measuring schedules and different distributions of trajectories $\mu(t)$, but for each of them this same control will apply. 

As an example, we can see that the control law at the start of the protocol ($t = 0$) recovers the known\cite{schmiedl2007optimal} initial trap displacement of $\Delta\lambda = [\tau/(t_f + 2\tau)]\Delta \mu$, where $\Delta \mu = \lambda_f - \mu(0)$. Assuming no further measurements are taken, the trajectory $\bar{\mu}(t)$ is given by the deterministic evolution $\tau \dot{\bar{\mu}} (t) = (\lambda^*(t) - \bar{\mu}(t))$:

\begin{equation}
\bar{\mu}(t) = \bar{\mu}(0) + \left( \frac{\lambda_f - \bar{\mu}(0)}{t_f + 2\tau} \right) t
\end{equation}

Thus, as $\Delta t \to 0$, the discrete-time POMDP framework recovers the constant-velocity trajectory and the discontinuous jump sizes characteristic of the Schmiedl-Seifert open-loop driving protocol\cite{schmiedl2007optimal}. In the case of initial measurement ($\bar{\mu}(0) = x_0$) the Abreu-Seifert protocol for fixed stiffness\cite{abreu2011extracting} is also recovered. For the case of periodic perfect measurements, the $\bar{\mu}(t)$ trajectory is a piecewise linear function with discontinuous jumps located at the measurement times, and the optimal control is a periodic ``reset'' of an open-loop control, consistent with \cite{garcia2025optimal,schuttler2025active}, see also Fig.~\ref{fig:trajectory}. 
The general control law, as derived within the POMDP framework, shows that the optimal agent chooses at each step the same microscopic control it would select if no further measurement were expected. This is also qualitatively consistent with the numerical approximations found by \cite{whitelam2023demon}.

\vspace{12pt}
\textbf{The Measurement Protocol and Finite-Time Trigger.} 
We can now tackle the problem of optimal measurement scheduling. The decision to measure reduces to the independent optimization of the informational cost $g_n(\Sigma^-)$ shown in Eq.~\ref{eq:decoupling_main}. 

First, we consider an instantaneous binary measurement $a_{obs} \in \{0, 1\}$, incurring a macroscopic cost $C$. While a perfect measurement of a continuous variable in principle requires infinite thermodynamic work\cite{sagawa2009minimal, parrondo2015thermodynamics}, $C$ here represents the macroscopic energy cost of activating a sensor of high enough resolution that any error is negligible relative to the physical length scales of the system dynamics. 

The optimal policy can be reduced, see Appendices A and C, to a piecewise dynamic programming minimum:

\begin{equation}\label{eq:piecewise_informational_cost}
\begin{split}
g_n(\Sigma^-) = \min_{a_{obs} \in \{0,1\}} \Big\{ &(1 - a_{obs}) g_{n-1}(\alpha^2\Sigma^- + \Sigma_{eq}) \\
&+ a_{obs} \left[ C - A_n\Sigma^- + g_{n-1}(\Sigma_{eq}) \right] \Big\}.
\end{split}
\end{equation}

Here $A_n = -P_n/2$ is the thermodynamic yield. Rearranging the minimization isolates the decision boundary for measurements:
\begin{equation}\label{eq:threshold_informational}
g_{n-1}(\alpha^2\Sigma^- + \Sigma_{eq}) - g_{n-1}(\Sigma_{eq}) \ge C - A_n \Sigma^-.
\end{equation}
The right side represents the net immediate cost of the measurement step: the observation cost $C$ reduced by the immediate profit $-A_n \Sigma^-$. The left side represents the difference in future informational costs for the different measurement choices. 


Since work can be extracted from reduction of uncertainty, the informational cost $g_n(\Sigma)$ is a decreasing function of $\Sigma$: a larger variance yields a better (more negative) cost (see the SM). The choice of measurement results from a compromise: Should the demon pay a cost $C$ to collapse the variance now and immediately harvest a gain $A_n \Sigma^-$, or allow the variance to grow for a larger potential gain in the future? 

The optimal demon delays measurement, waiting until the expected thermal fluctuations are large enough to reimburse both the absolute observation cost $C$ and the forfeited future thermodynamic yield, see Fig.~\ref{fig:trajectory}. Eq.~\ref{eq:threshold_informational} allows one to calculate \textit{backward}, from the trivial final step $n=0$, the exact threshold variance $\Sigma_{th}(n)$ above which an optimal agent will observe. Since the trajectory of the observed and unobserved thermal expansion is deterministic, the discrete sequence of optimal measurement times can be computed offline, proceeding \textit{forward} from the initial state $\Sigma_0=\Sigma_\kappa$, before the transport protocol begins.

\vspace{12pt}
\textbf{Deadline Devaluation.}
The value of information\cite{bauer2012efficiency} consists of extracting thermodynamic work from measured thermal fluctuations.
As discussed, the immediate available profit ($-A_n\Sigma$) is governed by the time-dependent coefficient $A_n=-P_n/2$ (see Eq.~\ref{eq:p_exact}).
As the deadline approaches ($n \to 1$), the thermodynamic yield decreases from $A_\infty = \kappa/2$ toward its terminal value $A_0 = 0$. The break-even variance threshold required to compensate the measurement cost, $C / A_n$, diverges. We term ``deadline-induced blindness''  the temporal region close to $t_f$, shaded in grey in Fig.~\ref{fig:trajectory}, where $\Sigma_{th}(n)$ exceeds the maximum physical variance of the thermal bath ($\Sigma_\kappa = k_B T / \kappa$). Since the bath cannot physically generate an expected fluctuation large enough to reimburse $C$, observations here are never profitable. 

Since $A_n$ monotonically decays with $n$, this blind region stretches backward from $t_f$ across the time horizon. As $C$ increases from $0$, the boundary of the blindness region expands. The limit $A_\infty = \kappa / 2$ sets a clear phase transition.  The maximum cost $C_{max} = \frac{1}{2}k_B T $ is a general threshold, beyond which the observation cost can never be repaid: The demon's optimal policy is to remain permanently blind. Conversely, costs $C < \frac{1}{2}k_B T$ can sustain an active steady-state limit cycle.

\begin{figure}
    \centering
    \includegraphics[width=0.9\linewidth]{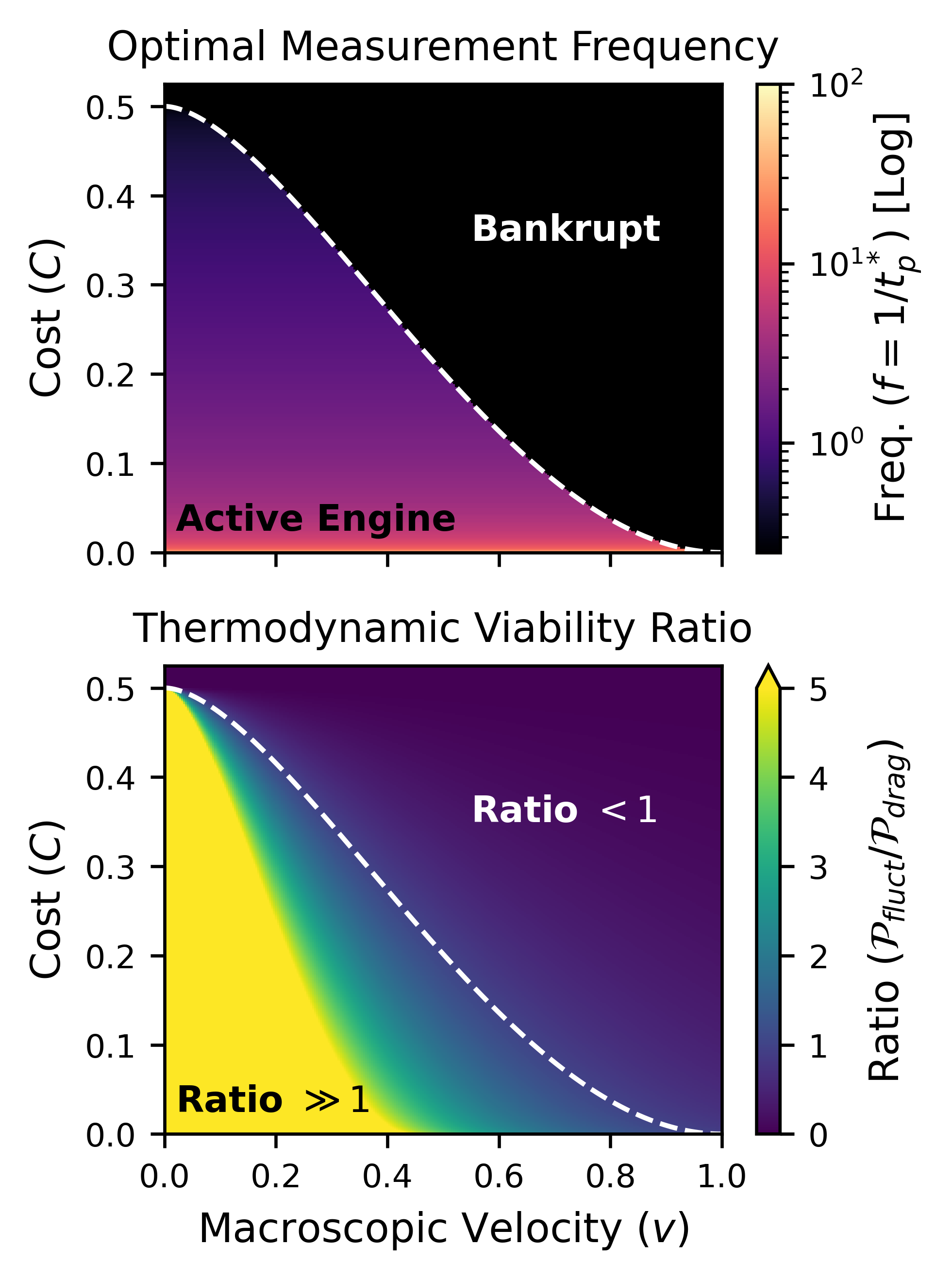}
\caption{\textbf{Thermodynamic phase space of the information engine with binary measurements at steady-state.} Heatmaps displaying the measurement frequency (top) and the thermodynamic viability ratio ($\mathcal{P}_{fluct}/\mathcal{P}_{drag}$) (bottom) as a function of macroscopic driving velocity $v$ and measurement cost $C$. The velocity envelope $C_{env}(v)$ (dashed line) separates the active, profitable engine regime from the net-dissipative regime. This operational space is bounded by two fundamental physical limits: the steady-state general threshold ($C = k_B T / 2$) at $v=0$, and the speed limit ($v = v_{max}$) at $C=0$, beyond which macroscopic viscous drag overwhelms the maximum extractable microscopic fluctuation power. Simulation parameters: $k_B T=1$, $\kappa=1$, $\gamma=1$.}
\label{fig:heatmaps}
\end{figure}

\vspace{12pt}
\textbf{Optimal Measurement Scheduling and the Lambert $W$ Transition.}
We consider now the steady-state regime, where both $\lambda_f \rightarrow \infty$ and $t_f \rightarrow \infty$ but at a finite macroscopic velocity $v = \lambda_f / t_f$. In practice, we can consider the limit $n\rightarrow \infty$ in Eq.~\ref{eq:lambda_exact}. In this regime the optimal trap placement reduces as expected to:

\begin{equation}   \label{eq:steady-state-control-discrete}
\lambda_k^* = \mu_k^+ + \frac{v \Delta t}{1 - \alpha}.
\end{equation}

The classical hydrodynamic lag\cite{schmiedl2007optimal} is recovered taking the limit $\Delta t\rightarrow 0$:

\begin{equation}   \label{eq:steady-state-control-continuous}
\lambda^*(t) = \mu(t) + \frac{\gamma v}{\kappa} = \mu(t) + \tau v.
\end{equation}

Again, this holds for any trajectory $\mu(t)$, irrespective to the history of measurements.

For the binary sensor the optimal measurement scheduling is periodic, since the variance threshold is fixed and the evolution of the variance is deterministic. To calculate the exact measurement period $\tau_{meas} = N^* \Delta t$ in the active steady-state regime, we can maximize the average thermodynamic profit rate of the demon. 
We obtain the following optimal continuous period (see the SM):
\begin{equation} \label{eq:lambert_time}
N^* = \frac{1 + \mathcal{W}_{-1}\left( -\frac{1}{e}\left(1 - \frac{C}{C_{max}}\right) \right)}{2 \ln \alpha},
\end{equation}

where $\mathcal{W}_{-1}$ denotes the lower branch of the Lambert $\mathcal{W}$ function\cite{corless1996lambert} and $\alpha = \exp(-\kappa \Delta t / \gamma)$. The mathematical singularity of the Lambert $W$ function ($N^* \to \infty$) corresponds to the physical threshold at $C_{max} = \frac{1}{2} k_B T$, while as one expects the period monotonically decrease as $C\rightarrow0$. 

As a technical side-note, the continuous-time Lambert $\mathcal{W}$ solution provides the analytical envelope for the optimal period, but the original discrete-time optimal policy cannot implement fractional intervals. Simply, the global Dynamic Programming algorithm must phase-lock to integer cycle lengths, $N_{cycle} \approx N^*$. 

\vspace{12pt}
\textbf{Macroscopic Profitability and the Engine Boundary.} 
We have proven that - in the steady-state regime - there is an optimal measurement frequency $1/N^*$ which depends exclusively on the observation cost $C$. This means that the optimal frequency is independent from the trap's macroscopic velocity $v = \lambda_f / t_f$.  On the other hand, the thermodynamics of the demon cannot be independent from $v$.

Indeed, for the system to operate as a net-positive information engine, the average microscopic power extracted from thermal fluctuations, $\mathcal{P}_{fluct} = [C_{max}(1 - \alpha^{2N^*}) - C] / (N^* \Delta t)$, must exceed the deterministic macroscopic power dissipated by viscous drag, $\mathcal{P}_{drag} = \gamma v^2$. We define the thermodynamic viability ratio as $\mathcal{P}_{fluct} / \mathcal{P}_{drag}$ and use it to find the bounded active sensing regime in the $(v, C)$ phase space. In the continuous-time limit $\Delta t\rightarrow 0$, we find the following profitability boundary (see the SM): 
\begin{align*}
C_{env}(v) = \\
\frac{1}{2} k_B T \left[ 1 - (v/v_{max})^2 + (v/v_{max})^2 \ln (v/v_{max})^2 \right],
\end{align*}
Where $v_{max} = \sqrt{\kappa k_B T / \gamma^2}$. Beyond this envelope, the demon bankrupts the engine, as it cannot operate at a net positive power with any measurement policy, as shown in Fig.~\ref{fig:heatmaps}. 

\vspace{12pt}
\textbf{The Variable-Precision Sensor.}
Generalizing the binary sensor to a continuous variable-precision mechanism, we assume the thermodynamic observation cost depends on its precision: $C(\Sigma^+, \Sigma^-) = c(1/\Sigma^+ - 1/\Sigma^-)$. The reduction of variance can be quantified by the Kalman gain\cite{kalman1960new} $L_k = \frac{\Sigma^- - \Sigma^+}{\Sigma^-}$.
Again, the optimal trap placement is the same as before, as it is fully independent from the measurement protocol and modality.
To obtain the optimal measuring policy, we have to minimizes the total informational cost:
\begin{equation}
\begin{split}
    g_n(\Sigma^-) = 
\min_{\Sigma^+ \le \Sigma^-} 
\Big[ c(1/\Sigma^+ - 1/\Sigma^-) -A_n(\Sigma^--\Sigma^+)\\
+ g_{n-1}(\alpha^2 \Sigma^+ + \Sigma_{eq}) \Big].
\end{split}    
\end{equation}

\begin{figure}
    \centering
    \includegraphics[width=1.0\linewidth]{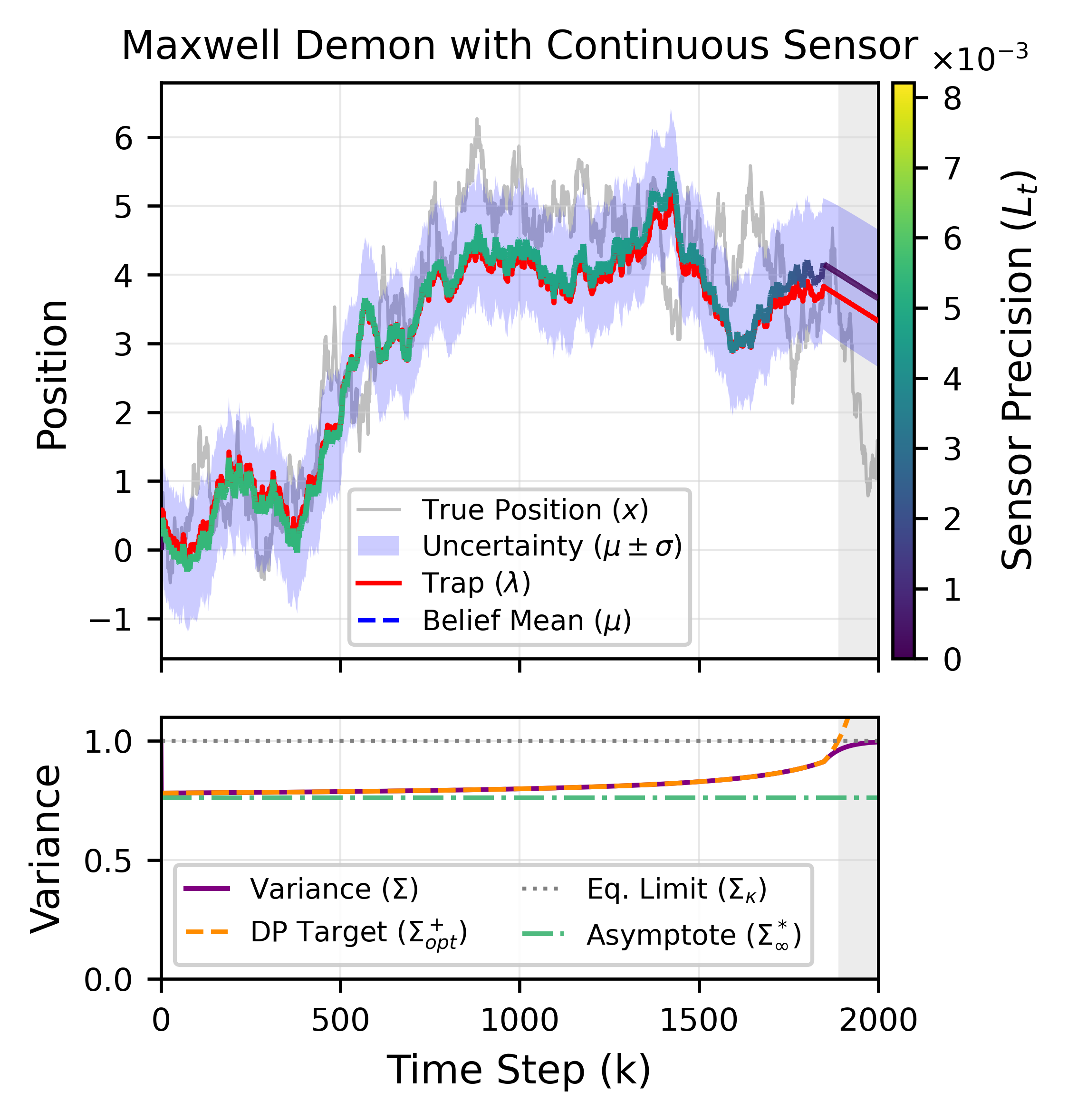}
\caption{\textbf{Optimal finite-time control dynamics of the information engine with variable-precision measurements.} (Top) Simulated spatial trajectory illustrating the true particle position ($x$, gray) and the optimal deterministic trap placement ($\lambda$, red). The agent's belief mean is color-coded by the instantaneous Kalman gain $L_k$, showing the active measurement intensity. The shaded blue region denotes the $1\sigma$ posterior uncertainty. (Bottom) Corresponding temporal evolution of the belief variance ($\Sigma$, purple). Throughout the trajectory, the demon continuously measures with just enough precision to keep the variance locked to the dynamic target $\Sigma_{opt}^+$ (dashed orange). As the deadline approaches, the target diverges triggering the terminal ``deadline blindness" phase (gray shading): the sensor is completely deactivated ($L_t \to 0$, visible as the dark shift in the belief mean colormap). Afterwards the variance converges to the thermal equilibrium variance $\Sigma_\kappa$ (dotted line). Simulation parameters: $k_B T=1$, $\kappa=1$, $\gamma=1$, $\Delta t=0.01$, $t_f=20$, $c=0.6 \, c_{max}$.}
\label{fig:thermostat}
\end{figure}

We derive in the SM that the optimal precision $\Sigma_{opt}^+(n)$ is independent from the prior variance $\Sigma^-$ at all times $n$. This means that the optimal demon either pays the cost to reduce $\Sigma^-$ to $\Sigma_{opt}^+(n)$ or, if $\Sigma^- < \Sigma_{opt}^+(n)$ it chooses not to use the sensor. In the finite-time regime, the spatial extraction coefficient $A_n$ monotonically decays as the deadline approaches ($n \to 0$). Again this deadline devaluation forces the required target $\Sigma_{opt}^+(n)$ to sweep upward. 

``Deadline blindness'' is triggered kinetically: the sensor fully deactivates the moment the growth of the target variance outpaces the forward thermal diffusion of the unmeasured system, i.e., when $\Sigma_{opt}^+(n-1) \ge \alpha^2 \Sigma_{opt}^+(n) + \Sigma_{eq}$. This can be seen in Fig.~\ref{fig:thermostat}, where the Kalman gain $L_k$ remains approximately stable for the majority of the trajectory but suddenly collapses close to the terminal point.  

Conversely, in the steady-state $n\rightarrow \infty$ and continuous-time $\Delta t\rightarrow \infty$ limits, the posterior variance locks into a constant that can be calculated in closed-form (see the SM):

\begin{equation}
\begin{split}
\Sigma_\infty^* &= \sqrt[3]{\frac{2ck_B T}{\kappa^2} + \sqrt{\frac{4c^2(k_B T)^2}{\kappa^4} + \frac{8c^3}{27\kappa^3}}} \\
&\quad + \sqrt[3]{\frac{2ck_B T}{\kappa^2} - \sqrt{\frac{4c^2(k_B T)^2}{\kappa^4} + \frac{8c^3}{27\kappa^3}}}.
\end{split}
\end{equation}

In this regime, unlike the cyclic regime of the binary sensor, the optimal policy is to operate as an ``information thermostat.'' Utilizing a variable-precision Kalman filter\cite{kalman1960new}, the demon constantly measures with a precision rate $\dot{\Sigma}_{meas}/\Sigma_\infty^{*2}$ (in the continuous-time limit) which exactly offset natural thermal diffusion, locking the posterior variance to the optimal target $\Sigma_\infty^*$. This balance requires $\dot{\Sigma}_{meas}=\frac{2\kappa}{\gamma}(\Sigma_{eq} - \Sigma_\infty^*)$.

\vspace{12pt}
\textbf{Macroscopic Profitability at the Steady State.} 
To derive the state space of the $(v,c)$ plane where the variable-precision sensor can act at a positive net power, we must again balance the optimal power extractable from fluctuations against the macroscopic viscous drag $\gamma v^2$. 
In the SM we derive this new envelope $c_{env}(v)$ for the steady state, in the continuous-time limit $\Delta t \rightarrow0$: 

\begin{equation}
c_{env}(v) = c_{max} \frac{(s^*(v/v_{max}))^3}{2 - s^*(v/v_{max})},
\end{equation}

Where $v_{max} = \sqrt{\kappa k_B T / \gamma^2}$ and the maximum cost threshold is $c_{max} = \frac{(k_B T)^2}{2\kappa}=\frac{k_B T}{2}\Sigma_\kappa$. The envelop is defined by $s^*(v/v_{max})$, which is the physically viable ($0< s \le 1$) root of the equation:
\begin{equation}
s^*(v/v_{max}) = \frac{4 - (v/v_{max})^2 - \sqrt{(v/v_{max})^4 + 8\nu}}{4}.
\end{equation} 

\begin{figure}
    \centering
    \includegraphics[width=1.0\linewidth]{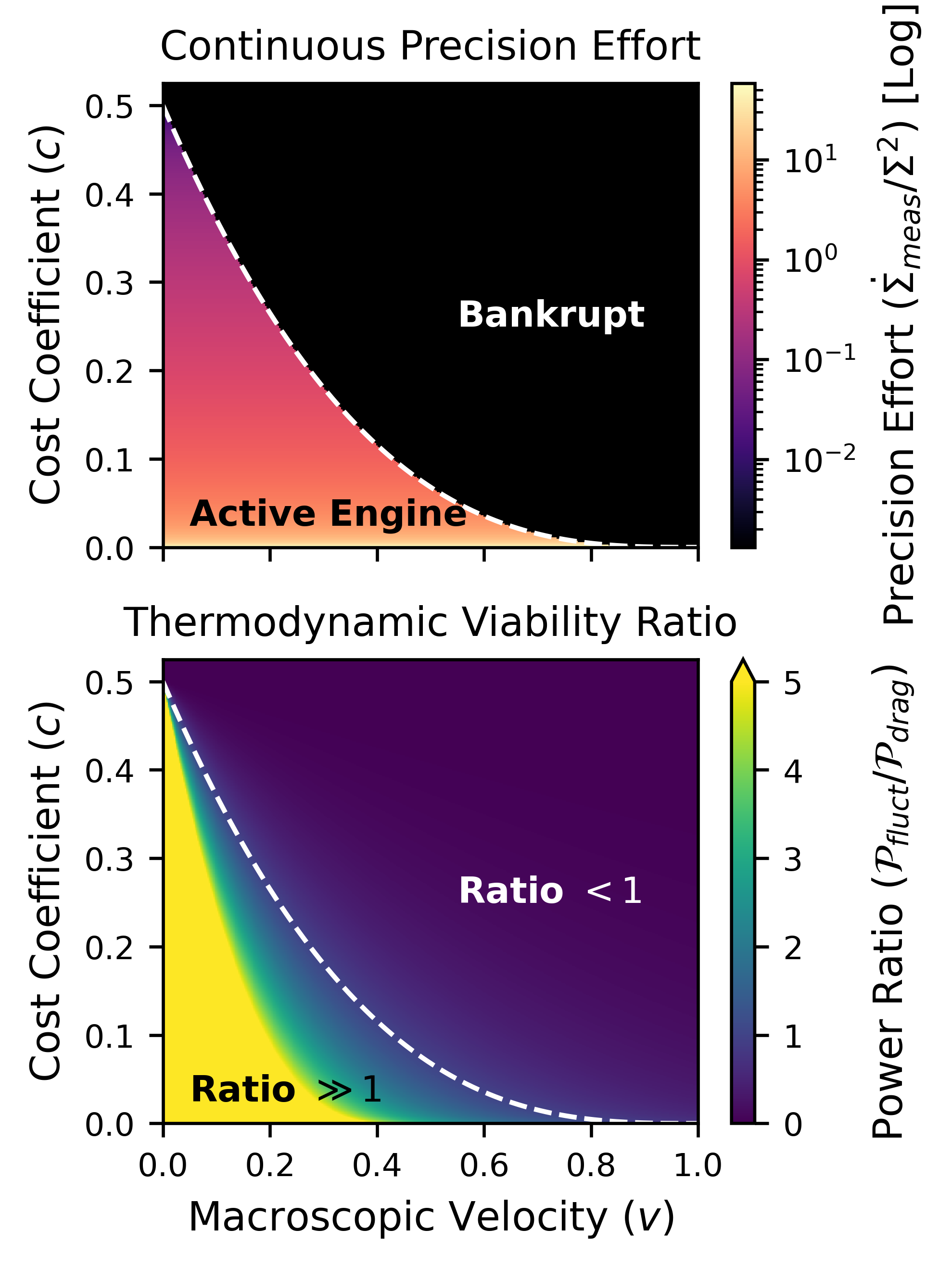}
\caption{\textbf{Thermodynamic phase space of the information engine with variable-precision measurements at steady-state.}. Heatmaps displaying the continuous precision rate ($\dot{\Sigma}_{meas}/\Sigma_\infty^{*2}$) (top) and the thermodynamic viability ratio ($\mathcal{P}_{fluct}/\mathcal{P}_{drag}$) (bottom) as a function of macroscopic driving velocity $v$ and the variable precision cost scalar $c$. The analytical envelope $c_{env}(v)$ (dashed line) separates the active, profitable engine regime from the net-dissipative regime. This operational space is bounded by two fundamental physical limits: the steady-state general threshold ($c = (k_B T)^2 / 2\kappa$) at $v=0$, and the speed limit ($v = v_{max}$) at $c=0$, beyond which macroscopic viscous drag overwhelms the maximum extractable continuous fluctuation power. Simulation parameters: $k_B T=1$, $\kappa=1$, $\gamma=1$.}
\label{fig:thermostat_phasespace}
\end{figure}

The velocity envelop shape can be seen as the dashed white line in Fig.~\ref{fig:thermostat_phasespace}. 
While both $C_{env}(v)$ and $c_{env}(v)$ share the same limiting velocity $v_{max}$, they differ in their asymptotic scaling near the stationary limit ($v \to 0$). The former approaches the maximum cost with a flat plateau ($-(v/v_{max})^2$), while for the continuous thermostat boundary the envelop drops linearly.

\vspace{12pt}
\textbf{Conclusion.}
In this work we apply control theory methods to information engines, tackling the control and measurement problem for an overdamped harmonic trap with costly sensors. Framing the problem as a discrete-time POMDP we are able to solve it analytically in closed-form. We show that the optimal feedback law recovers the continuous-time Schmiedl-Seifert limit, and generalizes it to any measurement scheduling. For we compute the optimal measurement protocols for two different realizations of sensors. In the case of a binary sensor, the optimal scheduling is a sequence of intermittent measurements, while the variable-precision sensor locks the variance onto an optimal, time-dependent, target.

Our analytical baselines can be used to gain insights on the finite-time boundaries of mesoscopic work extraction and the fundamental thermodynamic limits of an active information engine. Beyond the general cost thresholds and deadline-induced blindness, we calculate the macroscopic velocity envelopes $C_{env}(v)$ and $c_{env}(v)$ for the two sensors, which bound the regions of net-positive power output of the optimal agent.

The remarkable simplicity of our results derives from the one-dimensional nature of the Riccati recursive equation and the decoupling of the control and measurements tasks. These are direct consequences of the definition of the physical problem, i.e. the harmonicity of the trap and the Separation Principle
, guaranteed in the Linear-Quadratic-Gaussian framework within which we operate. 
While seemingly strict constraints, we note that these conditions should hold also for other, more general, physical setups such as the underdamped particle and the Active Ornstein-Uhlenbeck particles \cite{Bonilla_2019,garcia2025optimal,davis2026optimal,cocconi2023optimal,casert2024learning}, though with the increased complexity of having higher dimensional state representations.

On the contrary, severe deviations from this setup, such as anharmonic potentials or intrinsically non-Markovian fluctuations (e.g. a viscoelastic medium), will couple the measurement policy to the physical state and break our analytical solution. By setting the analytical boundaries of the uncoupled regime, our work helps define the baseline against which approximate numerical solvers and reinforcement learning algorithms must be compared to further advance the problem of optimal transport with costly measurements.

\vspace{12pt}

\textbf{Data and Code Availability}
The physical simulation code and plotting scripts required to reproduce the figures and thermodynamic boundaries presented in this manuscript are openly available via Zenodo at \url{https://doi.org/10.5281/zenodo.19485206}.

\begin{acknowledgments}
\textbf{Acknowledgments} The author also acknowledges Antonio Celani and Alessandro H. Ingrosso for fruitful discussions on control theory, and  Luis Reinalter and Clemens Bechinger for interesting discussions on the measurement scheduling problem.
\end{acknowledgments}
\textbf{Declaration of AI Use.} 
The author acknowledges the use of Gemini 3.1 Pro to assist with algebraic expansions, Python simulations, and LaTeX formatting. The author conceptualized the physical framework, verified all mathematical derivations and generated code, and assumes full responsibility for the originality, accuracy, and scientific integrity of this manuscript.

\bibliographystyle{apsrev4-2} 
\bibliography{bibtex}

\appendix

\section{A. Decoupling and Dimensional Reduction of the Cost Function}
\label{app:decoupling}

To distinguish the state transitions within a single step, we denote the prior state before measurement as $(\mu^-,\Sigma^-)$, the posterior state after measurement as $(\mu^+,\Sigma^+)$, and the prior state at the next step as $(\mu^{\prime -}, \Sigma^{\prime -})$.

In the following we demonstrate by induction that the POMDP cost function $V_n$ separates into an informational cost function $g_n(\Sigma)$ and a deterministic, quadratic physical cost $J_n(\mu, \lambda)$. We further show that the Linear-Quadratic-Gaussian (LQG) matrix for this spatial cost reduces to a 1D scalar, solution of a recurrence equation.

\subsection*{Base Case ($n=0$)}

At the deadline, the terminal spatial cost is defined by the expected work of the constrained final jump to the target $\lambda_f$. Evaluating this expectation cancels the spatial variance terms:
\begin{equation}\label{eq:V0}
V_0(\mu^-, \lambda) = \kappa\mu^-\lambda - \frac{1}{2}\kappa\lambda^2 - \kappa\lambda_f\mu^- + \frac{1}{2}\kappa\lambda_f^2.
\end{equation}
Because this terminal cost is independent of the posterior variance $\Sigma^+$, the value function decouples into $g_0(\Sigma^-) = 0$ and $J_0(\mu^-, \lambda) = V_0(\mu^-, \lambda)$. 

\subsection*{Inductive Step and Decoupling}

Assume at step $n-1$ the cost function is decoupled: $V_{n-1}(\mu^-, \Sigma^-, \lambda) = J_{n-1}(\mu^-, \lambda) + g_{n-1}(\Sigma^-)$. At step $n$, the Bellman optimality equation evaluates the measurement and control actions. Using $\lambda$ for the historical trap position and $\lambda'$ for the current action:
\begin{equation}\label{eq:bellman_full}
\begin{split}
V_n(\mu^-, \Sigma^-, \lambda) &= \min_{\Sigma^+} \Big\{ C(\Sigma^+, \Sigma^-) + \\
&\quad \mathbb{E}_{\mu^+} \left[ \min_{\lambda'} \left\{ \mathbb{E}_{x}[W] + V_{n-1}(\mu^{\prime-}, \Sigma^{\prime-}, \lambda') \right\} \right] \Big\}.
\end{split}
\end{equation}

Substituting the decoupled cost-to-go function gives:
\begin{equation}\label{eq:bellman_expanded}
\begin{split}
V_n(\mu^-, &\Sigma^-, \lambda) = \min_{\Sigma^+} \Big\{ C(\Sigma^+, \Sigma^-) + \\
 \mathbb{E}_{\mu^+} &\left[ \min_{\lambda'} \left\{ \mathbb{E}_{x}[W] + J_{n-1}(\mu^{\prime-}, \lambda') + g_{n-1}(\Sigma^{\prime-}) \right\} \right] \Big\}.
\end{split}
\end{equation}

To separate the inner minimization over the physical action $\lambda'$ from the informational cost, we rely on three physical independencies:
\begin{enumerate}
    \item The expected thermodynamic work of instantaneously translating a harmonic trap is independent of the spatial variance. Since the trap stiffness $\kappa$ is constant, the $\mathbb{E}[x^2]$ terms cancel exactly: $\mathbb{E}_x[W] = -\kappa\mu^+\lambda' + \frac{1}{2}\kappa(\lambda')^2 + \kappa\mu^+\lambda - \frac{1}{2}\kappa\lambda^2$.
    \item The transition of the prior mean is deterministic and invariant to the measurement uncertainty $\Sigma^+$: $\mu^{\prime-} = \alpha \mu^+ + (1 - \alpha) \lambda'$.
    \item The information dynamics $\Sigma^{\prime-} = \alpha^2 \Sigma^+ + \Sigma_{eq}$ are  deterministic and independent of the physical trap placement $\lambda^{\prime}$.
\end{enumerate}

Since $\Sigma^{\prime-}$ is independent of $\lambda'$, the future informational cost $g_{n-1}(\Sigma^{\prime-})$ acts as a constant with respect to the inner minimization over $\lambda'$ and factors out. We define the remaining inner minimization as the intermediate physical cost $\tilde{J}_n(\mu^+, \lambda)$:
\begin{equation}\label{eq:J_tilde}
\tilde{J}_n(\mu^+, \lambda) = \min_{\lambda'} \left\{ \mathbb{E}_{x}[W] + J_{n-1}(\alpha \mu^+ + (1 - \alpha) \lambda', \lambda') \right\}.
\end{equation}

The system is Linear-Quadratic, so $\tilde{J}_n$ is quadratic in the posterior mean $\mu^+$, possessing a spatial curvature defined as $P_{xx,n} \equiv \partial^2 \tilde{J}_n / \partial (\mu^+)^2$. 

We can evaluate the outer expectation over the stochastic measurement outcome $\mu^+$, which is distributed as $\mathcal{N}(\mu^-, \Sigma^- - \Sigma^+)$. Applying the identity $\mathbb{E}[X^2] = \mathbb{E}[X]^2 + \text{Var}(X)$ to the quadratic intermediate cost $\tilde{J}_n$ gives:
\begin{equation}\label{eq:expectation_mu_plus}
\mathbb{E}_{\mu^+} [ \tilde{J}_n(\mu^+, \lambda) ] = \tilde{J}_n(\mu^-, \lambda) + \frac{1}{2}P_{xx,n}(\Sigma^- - \Sigma^+).
\end{equation}

We define the physical cost evaluated at the prior mean as $J_n(\mu^-, \lambda) \equiv \tilde{J}_n(\mu^-, \lambda)$. Substituting this expansion back into the full Bellman equation separates the state space into independent spatial and informational recurrences:

\begin{equation}\label{eq:bellman_decoupled}
\begin{split}
V_n(\mu^-, \Sigma^-, \lambda) = &J_n(\mu^-, \lambda) + \\
&\min_{\Sigma^+} \Big\{ C(\Sigma^+, \Sigma^-) + \frac{1}{2}P_{xx,n}(\Sigma^- - \Sigma^+) +\\
&\quad\quad\quad g_{n-1}(\alpha^2 \Sigma^+ + \Sigma_{eq}) \Big\}.
\end{split}
\end{equation}
The minimization over $\Sigma^+$ defines the current informational cost $g_n(\Sigma^-)$, confirming the inductive hypothesis.

\subsection*{Riccati Dimensional Reduction}
Standard discrete-time LQG control requires tracking an augmented joint state space $\mathbf{z} = [\mu^+, \lambda]^T$ and solving a $2 \times 2$ Riccati cost matrix $\mathbf{P}_n$. 

Evaluating the stationary point $\partial / \partial \lambda' (\cdot)= 0$ inside the intermediate cost $\tilde{J}_n$ (Eq.~\ref{eq:J_tilde})  removes the previous trap position $\lambda$, as the expected work contains no cross-coupling term $\lambda'\lambda$. The optimal control law $\lambda^{\prime*}(\mu^+)$ is therefore independent of the previous trap position.

Substituting $\lambda^{\prime*}$ back into the objective function allows $\lambda$ to exit the minimization operator. Mapped to the standard $2 \times 2$ augmented LQG formulation, the off-diagonal coupling term is constant at $P_{x\lambda, n} = \kappa$, and similarly $P_{\lambda\lambda, n} = -\kappa$ for all finite $n$. Thus, the multi-variable Riccati matrix equation is unnecessary; the spatial dynamics are captured entirely by the scalar recurrence of the spatial curvature $P_{xx, n}$, which for simplicity we can rename, $P_{xx, n} \equiv P_n$.

\section{B. Analytical Solution of the Discrete Riccati Recurrence}
\label{app:riccati}

The dimensional reduction of the POMDP reduces the spatial cost to a single scalar recurrence\cite{bechhoefer2021control}:
\begin{equation}\label{eq:riccati_scalar}
P_{n} = P_{n-1}\alpha^{2} - \frac{\Gamma^{2}}{\Omega},
\end{equation}
where $\Gamma = (1-\alpha)(P_{n-1}\alpha - \kappa)$ and $\Omega = (1-\alpha)(P_{n-1}(1-\alpha) + 2\kappa)$.

Substituting these explicit coefficients into the recurrence, expanding the algebraic fraction, and grouping terms by $P_{n-1}$ reduces the non-linear equation to a M\"obius transformation:
\begin{equation}\label{eq:moebius}
P_n = \frac{2\alpha\kappa P_{n-1} - (1-\alpha)\kappa^2}{(1-\alpha)P_{n-1} + 2\kappa}
\end{equation}

This recurrence be linearized via the substitution:

\begin{equation}\label{eq:substitution}
y_n = \frac{1}{P_n + \kappa}.
\end{equation}

Inserting this substitution into Eq.~\ref{eq:moebius} transforms it into a simpler arithmetic progression:
\begin{equation}\label{eq:progression}
y_n = y_{n-1} + \frac{1-\alpha}{\kappa(1+\alpha)}.
\end{equation}

For a finite-time transport protocol, the cancellation of the quadratic terms in the terminal cost dictates the boundary condition $P_0 = 0$, which forces the terminal sequence value to $y_0 = 1/\kappa$. Evaluating the arithmetic progression backwards from this deadline yields:
\begin{equation}\label{eq:yn_solved}
y_n = \frac{1}{\kappa} + n\frac{1-\alpha}{\kappa(1+\alpha)}.
\end{equation}

Inverting the initial substitution via $P_n = \frac{1}{y_n} - \kappa$ provides the closed-form analytical solution provided in the main text for the dynamic coefficient at any $n$:

\begin{equation}\label{eq:Pn_solved}
P_n = -\kappa \left[ \frac{n(1-\alpha)}{1+\alpha+n(1-\alpha)} \right].
\end{equation}

\section{C. Measurement Policy}
\label{app:measurement}

\subsection*{The Informational Cost Function}
As derived in SM.A, the mathematical decoupling of the Bellman equation isolates the informational cost into the standalone minimization shown in Eq.~\ref{eq:bellman_decoupled}. The average over the posterior mean - $\mathbb{E}_{\mu^+}[\cdot]$ - shown in Eq.~\ref{eq:expectation_mu_plus} brings out the additional term $\frac{1}{2}P_n(\Sigma^- - \Sigma^+)$. 

By defining the thermodynamic energy bonus as $A_n = -P_n / 2$, this term becomes:
\begin{equation}\label{eq:energy_bonus}
-A_n(\Sigma^- - \Sigma^+).
\end{equation}

One can think of this negative cost as the immediate spatial energy bonus extracted by localizing the particle prior to executing the trap displacement. The Bellman equation for the informational cost reduces to minimizing the sum of the observation cost, this immediate spatial energy bonus, and the expected future informational cost:
\begin{equation}\label{eq:gn_min}
\begin{split} 
g_n(\Sigma^-) = \min_{\Sigma^+} \Big[ & C(\Sigma^+, \Sigma^-) \\ 
& - A_n(\Sigma^- - \Sigma^+) + g_{n-1}(\alpha^2 \Sigma^+ + \Sigma_{eq}) \Big]. 
\end{split}
\end{equation}

\subsection*{The Piecewise Minimum For The Binary Sensor}
For the binary measurement, the minimization separates into the two branches of Eq.~6:
\begin{equation}\label{eq:gn_binary}
\begin{split}
g_n(\Sigma^-) = \min_{a_{obs} \in \{0,1\}} \Big\{ &(1-a_{obs})g_{n-1}(\alpha^2\Sigma^- + \Sigma_{eq}) \\
&+ a_{obs}[C - A_n\Sigma^- + g_{n-1}(\Sigma_{eq})] \Big\}.
\end{split}
\end{equation}
In the measured branch ($a_{obs}=1$), the residual variance collapses, leaving only the thermal diffusion over a single step, $\Sigma_{eq}$. The trigger condition dictates that the controller measures if and only if:
\begin{equation}\label{eq:trigger}
g_{n-1}(\alpha^2\Sigma^- + \Sigma_{eq}) - g_{n-1}(\Sigma_{eq}) \ge C - A_n\Sigma^-.
\end{equation}

In a thermodynamic harmonic engine, variance represents extractable potential energy. The informational cost function $g_n(\Sigma)$ can be proven to be monotonically decreasing in $\Sigma$: a larger variance yields a better (more negative) cost. 

\vspace{12pt}
\textbf{Proof of Monotonicity:} We prove by induction that $\partial g_n / \partial \Sigma \le 0$.
\begin{itemize}
    \item \textbf{Base Case ($n=0$):} At the terminal deadline, there is no future extraction, so $g_0(\Sigma) = 0$. The derivative is therefore zero.
    \item \textbf{Step 1 ($n=1$):} Evaluating the Bellman equation one step before the deadline yields $g_1(\Sigma) = \min\{0, C - A_1\Sigma\}$. Because the physical extraction coefficient $A_1 > 0$, the measured branch decreases with $\Sigma$. Thus, $\partial g_1 / \partial \Sigma \le 0$.
    \item \textbf{Inductive Step ($n$):} Assume $g_{n-1}(\Sigma)$ is monotonically decreasing ($\partial g_{n-1} / \partial \Sigma \le 0$). We evaluate the derivatives of both branches of $g_n(\Sigma)$. For the unmeasured branch, $\frac{\partial}{\partial\Sigma}g_{n-1}(\alpha^2\Sigma + \Sigma_{eq}) = \alpha^2 \frac{\partial g_{n-1}}{\partial\Sigma} \le 0$. For the measured branch, $\frac{\partial}{\partial\Sigma}[C - A_n\Sigma + g_{n-1}(\Sigma_{eq})] = -A_n < 0$. Because both branches are non-increasing functions of $\Sigma$, $g_n(\Sigma)$ must be monotonically decreasing.
\end{itemize}

\section{D. Macroscopic Profitability Boundary for Perfect Measurement}
\label{app:profitability}

We derive the profitability boundary $C_{env}(v)$ for an information engine with a perfect binary sensor. We evaluate the continuous-time limits where the maximum achievable fluctuation power is limited by the macroscopic viscous drag.

\subsection*{The Continuous-Time Extracted Power}
In the discrete-time formulation, the average microscopic power extracted from thermal fluctuations is $\mathcal{P}_{fluct} = [C_{max}(1 - \alpha^{2N^*}) - C] / (N^* \Delta t)$, where $C_{max} = \frac{1}{2} k_B T$. To evaluate the continuous-time measurement period $t_p = N^* \Delta t$ we can simply write explicitly the relaxation parameter $\alpha = \exp(-\kappa \Delta t / \gamma)$. This yields the power equation:
\begin{equation}\label{eq:P_fluct}
\mathcal{P}_{fluct}(t_p) = \frac{C_{max} (1 - e^{-2\kappa t_p / \gamma}) - C}{t_p}
\end{equation}

\subsection*{The Optimality Condition}
For the system to operate at the active engine boundary, it must maximize its average thermodynamic profit rate with respect to the measurement period. We enforce the stationarity condition $\frac{\partial \mathcal{P}_{fluct}}{\partial t_p} = 0$:
\begin{equation}\label{eq:dP_dt}
\frac{\partial}{\partial t_p} \left[ \frac{C_{max} (1 - e^{-2\kappa t_p / \gamma}) - C}{t_p} \right] = 0
\end{equation}
Defining the dimensionless parameter $x = 2\kappa t_p / \gamma$, which represents the ratio of the measurement period to the trap relaxation time, the derivative requires:
\begin{equation}\label{eq:optimality_x}
C_{max} e^{-x} (x + 1) = C_{max} - C
\end{equation}
Isolating the required thermodynamic measurement cost $C$ yields:
\begin{equation}\label{eq:C_req}
C = C_{max} [1 - e^{-x}(x + 1)]
\end{equation}

Solving for the optimal measurement period $t_p^*$, we use the known inversion of the equation $Y e^Y = Z$. Rearranging Eq.~ \ref{eq:C_req}, we have:

\begin{equation}
-(x + 1)e^{-(x + 1)} = -\frac{1}{e}\left( 1 - \frac{C}{C_{max}} \right).
\end{equation}

This maps to the defining equation of the Lambert $W$ function, $W(z)e^{W(z)} = z$ \cite{corless1996lambert}. Applying the function to both sides gives:

\begin{equation}
-(x + 1) = W\left( -\frac{1}{e}\left( 1 - \frac{C}{C_{max}} \right) \right).
\end{equation}

Since the physical time ratio $x = 2\kappa t_p / \gamma $ must be positive ($x > 0$), the argument $-(x + 1)$ is less than $-1$. The Lambert $W$ function is double-valued on the interval $[-1/e, 0)$\cite{corless1996lambert} for which $C_{max}>C>0$, but the principal branch $W_0(z) \ge -1$ yields non-physical negative times $x \le 0$. The solution is found by the lower real branch, $W_{-1}(z) \le -1$. Solving for $x$:

\begin{equation}-(x + 1) = W_{-1}\left( -\frac{1}{e}\left( 1 - \frac{C}{C_{max}} \right) \right).\end{equation}

and substituting the physical constants back into the parameter yields Eq. \ref{eq:lambert_time}

\subsection*{Maximized Extracted Power}
Substituting the optimality constraint calculated in Eq. \ref{eq:C_req} back into the power equation defines the maximum theoretically extractable power for a given dimensionless period $x$:
\begin{equation}\label{eq:Pmax_sub}
\mathcal{P}_{max}(x) = \frac{2\kappa}{\gamma x} \left[ C_{max}(1 - e^{-x}) - C_{max}(1 - e^{-x} - x e^{-x}) \right]
\end{equation}
\begin{equation}\label{eq:Pmax_final}
\mathcal{P}_{max}(x) = \frac{2\kappa C_{max}}{\gamma} e^{-x}
\end{equation}

\subsection*{The Speed Limit}
The physical speed limit of the engine occurs when the observation cost is zero Here, the optimal policy dictates measuring each step, driving the period $t_p \to 0$, and therefore $x \to 0$. In this limit, the limiting factor is the macroscopic drag:
\begin{equation}\label{eq:speed_limit_bal}
\gamma v_{max}^2 = \frac{2\kappa C_{max}}{\gamma}
\end{equation}
Solving for the limiting velocity yields:
\begin{equation}\label{eq:vmax}
v_{max} = \sqrt{\frac{2\kappa C_{max}}{\gamma^2}} = \sqrt{\frac{\kappa k_B T}{\gamma^2}}.
\end{equation}

\subsection*{The $C_{env}(v)$ Envelope}

At the profitability boundary, the optimized fluctuation power equals the power dissipated by viscous drag:
\begin{equation}\label{eq:boundary_power}
\mathcal{P}_{max}(x) = \gamma v_{max}^2 e^{-x} = \gamma v^2.
\end{equation}
Solving this equality for the dimensionless period $x$ yields identities for the boundary constraints:
\begin{equation}\label{eq:x_identities}
e^{-x} = \left( \frac{v}{v_{max}} \right)^2 \implies x = -\ln \left( \frac{v}{v_{max}} \right)^2.
\end{equation}
Substituting these identities back into the stationary cost constraint (Eq.~\ref{eq:C_req}) allows us to obtain the velocity envelope:
\begin{equation}\label{eq:Ccrit_sub}
C_{env}(v) = C_{max} \left[ 1 - \left(\frac{v}{v_{max}}\right)^2 \left( -\ln \left( \frac{v}{v_{max}} \right)^2 + 1 \right) \right].
\end{equation}

\section{E. Variable-Precision Sensor}
\label{app:thermostat_limits}

In this section we derive the optimal measurement policy for the variable-precision sensor, both for the finite-time horizon and the steady-state. We evaluate $\Sigma_\infty^*$ and the profitability boundary $c_{env}(v)$ in the continuous-time limit ($\Delta t \to 0$).

\subsection{Finite-Time Optimal Measurement Policy}
\label{app:thermostat}

For a variable-precision sensor, the measurement cost is inversely proportional to the resolved spatial uncertainty: $C(\Sigma^+, \Sigma^-) = c(1/\Sigma^+ - 1/\Sigma^-)$. To determine the optimal measurement policy at step $n$, the agent minimizes the total informational cost:

\begin{equation}\label{eq:gn_continuous}
\begin{split}
g_n(\Sigma^-) = \min_{\Sigma^+ \le \Sigma^-} \Big[ &c\left(\frac{1}{\Sigma^+} - \frac{1}{\Sigma^-}\right) - A_n(\Sigma^- - \Sigma^+) \\
&+ g_{n-1}(\alpha^2 \Sigma^+ + \Sigma_{eq}) \Big].
\end{split}
\end{equation}

To find the minimum, we differentiate the bracketed objective function with respect to the measurement outcome $\Sigma^+$ and set it to zero:
\begin{equation}\label{eq:stationarity}
-\frac{c}{(\Sigma^+)^2} + A_n + \alpha^2 g'_{n-1}(\alpha^2 \Sigma^+ + \Sigma_{eq}) = 0.
\end{equation}

Note that the terms containing the prior variance $\Sigma^-$, being constants during the differentiation over $\Sigma^+$, have vanished from the stationarity condition: The optimal measurement depends exclusively on the system parameters ($c, \alpha, \Sigma_{eq}$), the physical extraction coefficient ($A_n$), and the future value function gradient ($g'_{n-1}$). 

We define this optimal root as the dynamic target precision, $\Sigma_{opt}^+(n)$. Since it is independent of the prior state $\Sigma^-$, the optimal control policy reduces to a simple binary choice:

\begin{enumerate}
    \item \textbf{Measurement Regime ($\Sigma^- > \Sigma_{opt}^+(n)$):} If thermal diffusion has pushed the prior spatial uncertainty above the optimal setpoint, the agent injects the measurement energy required to reduce the posterior variance down to the target: $\Sigma^+ = \Sigma_{opt}^+(n)$.
    \item \textbf{Idle Regime ($\Sigma^- \le \Sigma_{opt}^+(n)$):} If the prior uncertainty is already at or below the optimal setpoint, measuring incurs a cost without yielding a sufficient energetic return. The agent leaves the state unaltered: $\Sigma^+ = \Sigma^-$.
\end{enumerate}

\subsection{Steady State Net Power and Optimal Precision}
\label{app:profit_limit}

For the variable-precision sensor, the measurement cost scales as $C(\Sigma^+, \Sigma^-) = c (1/\Sigma^+ - 1/\Sigma^-)$. 
To derive the continuous-time thermodynamic profit rate from the discrete-time control framework, we evaluate the discrete profit generated during a single interval $\Delta t$ and take the asymptotic limit $\Delta t \to 0$.

Let us consider a steady state, where we have locked into a given precision $\Sigma_{ss}$. Each step, the 
 variance evolves via thermal diffusion for $\Delta t$ to $\Sigma'=\alpha^2 \Sigma_{ss} + (1-\alpha^2)\Sigma_{eq}$, gaining a (fixed) amount $\Delta \Sigma = (1-\alpha^2)(\Sigma_{eq}-\Sigma_{ss})$ and paying a cost of $c (1/\Sigma_{ss} - 1/\Sigma')$

\subsection{The Thermodynamic Profit for a Discrete Step}

The net thermodynamic profit generated during a single discrete step is the spatial energy bonus minus the observation cost:
\begin{equation}
    \Delta \mathcal{P} = A_\infty  \Delta \Sigma - c \left( \frac{1}{\Sigma_{ss}} - \frac{1}{\Sigma'} \right).
\end{equation}
Factoring out $\Delta \Sigma$ from the cost term gives the discrete profit per step:
\begin{equation}
\Delta \mathcal{P} = \Delta \Sigma \left[ A_\infty - \frac{c}{\Sigma_{ss} \Sigma'} \right].
\end{equation}

\subsection{The Continuous Limit ($\Delta t \to 0$)}
The continuous profit rate is defined as the limit of the discrete profit over the timestep: $\mathcal{P}_{fluct} = \lim_{\Delta t \to 0} \frac{\Delta \mathcal{P}}{\Delta t}$.

To evaluate this, we apply the small $\Delta t$ limits to the discrete components. The relaxation parameter gives $1 - \alpha^2 \approx \frac{2\kappa}{\gamma} \Delta t$ and the variance rate $\lim_{\Delta t \to 0} \frac{\Delta \Sigma}{\Delta t} = \frac{2\kappa}{\gamma} (\Sigma_{eq} - \Sigma)$. Furthermore we have that $A_\infty = \frac{\kappa}{2}$ and, at the denominator, $\Sigma_{ss}\Sigma' =\Sigma_{ss}^2$

Substituting these into the profit rate equation gives:

\begin{equation}\label{eq:power_continuous}
\begin{split}
\mathcal{P}_{fluct}(\Sigma_{ss}) &= \lim_{\Delta t \to 0} \frac{\Delta \Sigma}{\Delta t} \left[ A_\infty - \frac{c}{\Sigma_{ss} \Sigma^-} \right] \\
&= \frac{2\kappa}{\gamma}(\Sigma_{eq} - \Sigma_{ss}) \left[ \frac{\kappa}{2} - \frac{c}{\Sigma_{ss}^2} \right].
\end{split}
\end{equation}

The optimal $\Sigma_\infty^*$ satisfies the stationarity condition ${\partial \mathcal{P}_{fluct}}/{\partial \Sigma_{ss}}|_{\Sigma_\infty^*} = 0$:

\begin{equation}\label{eq:optimality_stationary}
\frac{c}{\Sigma_\infty^{*2}} \left( \frac{2k_B T}{\kappa \Sigma_\infty^*} - 1 \right) = \frac{\kappa}{2}
\end{equation}

Multiplying by $2\kappa \Sigma_\infty^{*3}$ reorganizes this into a cubic polynomial:

\begin{equation}
\kappa^2\Sigma_\infty^{*3} + 2c\kappa\Sigma_\infty^* - 4ck_BT = 0
\end{equation}

Since the physical parameters are positive, the discriminant is positive, guaranteeing a single real root. Applying Cardano's formula returns the expression for the optimal continuous precision $\Sigma_\infty^*$ shown in the main text:

\begin{equation}
\begin{split}
\Sigma_\infty^* &= \sqrt[3]{\frac{2ck_B T}{\kappa^2} + \sqrt{\frac{4c^2(k_B T)^2}{\kappa^4} + \frac{8c^3}{27\kappa^3}}} \\
&\quad + \sqrt[3]{\frac{2ck_B T}{\kappa^2} - \sqrt{\frac{4c^2(k_B T)^2}{\kappa^4} + \frac{8c^3}{27\kappa^3}}}.
\end{split}
\end{equation}

\subsection{Dimensionless Formulation}

To evaluate the macroscopic transport limits, we define the dimensionless variance $s = \kappa \Sigma_\infty^* / k_B T$, constrained to $s \in (0, 1]$. We can thus rewrite the optimality condition of Eq.~\ref{eq:optimality_stationary} to isolate the measurement cost $c$:

\begin{equation}
c = \frac{(k_B T)^2}{2\kappa} \frac{s^3}{2-s}
\end{equation}

The absolute limit occurs at the boundary of net-zero power ($s \to 1$), defining the critical sensor cost:

\begin{equation}
c_{max} = \frac{(k_B T)^2}{2\kappa}
\end{equation}

The required operating cost for any optimal dimensionless variance $s$ therefore reduces to $c(s) = c_{max} \frac{s^3}{2-s}$.

\vspace{12pt}
\subsection{The Envelope $c_{env}(v)$}

Substituting $c(s)$ back into the continuous power equation of Eq.~\ref{eq:power_continuous}, we get the optimal net power as a function of the dimensionless variance:

\begin{equation}
\mathcal{P}_{fluct}(s) = \frac{2\kappa k_B T}{\gamma} \frac{(1 - s)^2}{2 - s}
\end{equation}

To find the operational envelope for any arbitrary macroscopic velocity $v$, we equate the optimal net power to the required drag dissipation:

\begin{equation}
\gamma v^2 = \gamma v_{max}^2 \frac{2(1-s)^2}{2-s}
\end{equation}

Defining the squared velocity ratio $\nu = (v/v_{max})^2$, this expands into a quadratic condition for the required steady-state variance:

\begin{equation}
2s^2 + (\nu - 4)s + 2(1 - \nu) = 0
\end{equation}

Taking the physical root ($s \le 1$) yields the explicit spatial threshold required to sustain velocity $v$:

\begin{equation}
s^*(v) = \frac{4 - \nu - \sqrt{\nu^2 + 8\nu}}{4}
\end{equation}

Substituting this dynamic root back into the cost function closes the system, providing the profitability boundary for the variable-precision engine:

\begin{equation}
c_{env}(v) = c_{max} \frac{(s^*(v))^3}{2 - s^*(v)}
\end{equation}

\end{document}